\documentclass[graphicx,amsmath,amssymb,pra,twocolumn]{revtex4-1}

\usepackage{graphicx}		
\usepackage{color}
\usepackage{amsmath}
\usepackage{subfigure}
\usepackage{upgreek}
\usepackage{tabularx}

\usepackage{dsfont}

\begin{document}
	
	\title{Self-organized transport in noisy dynamic networks}	
	\author{Frederic Folz$^1$}
	\author{Kurt Mehlhorn$^2$}
	\author{Giovanna Morigi$^1$}
	\affiliation{$^1$Theoretische Physik, Universit\"at des Saarlandes, 66123 Saarbr\"ucken, Germany \\ $^2$Algorithms and Complexity Group, Max-Planck-Institut f\"ur Informatik, Saarland Informatics Campus, 66123 Saarbr\"ucken, Germany}
	
	\date{\today}
	
	\begin{abstract} 
We present a numerical study of multi-commodity transport in a noisy, nonlinear network. The nonlinearity determines the dynamics of the 
edge capacities, which can be 
amplified or suppressed depending on the local current flowing across 
an edge. We consider network self-organization for three different nonlinear functions: For all three 
we identify parameter regimes where noise leads to self-organization into {more} robust topologies, that are 
not found by the sole noiseless dynamics. 
Moreover, the interplay between noise and specific functional behavior of the nonlinearity
gives rise to different features, such as (i) continuous or discontinuous responses to the demand strength and (ii) either single or multi-stable solutions. Our study shows the crucial role of the activation function on noise-assisted phenomena. 
\end{abstract}
	
\maketitle

\section{Introduction}
\label{Sec:introduction}

Networks are a commonly used concept in many disciplines and powerful models for transport. Efficient routing of commodities such as water, power, or information from sources to sinks can be described as a problem of connecting nodes on a graph for given constraints and requirements \cite{Yang:1998, Feremans:2003, Farahani:2013}. There are different approaches to network design. 
{Some} consist of solving differential equations, which are derived from an appropriately identified cost function \cite{Yeung:2009,Banavar:2000}. The chosen rule determines the dynamics of the edges connecting the nodes, whose stable fixed point is a target topology. Extensive studies show how, for a given class of power law functions determining the equations of motion, a variation of the exponent can give rise to phase transitions in the network structure, from spanning trees that minimize the cost to loops that maximize the robustness \cite{Banavar:2000,Lonardi:2021, Lonardi:2022}.

Another approach is based on bio-inspired algorithms. 
Prominent examples are algorithms inspired by the structures formed by ant colonies \cite{Dorigo:2006} or by the filaments of Physarum polycephalum, a single-celled organism that is also known as true slime mold \cite{Oettmeier:2020}. Despite its lack of any form of a nervous system, Physarum polycephalum is able to find good solutions to small instances of popular optimization problems, such as finding the shortest path through a maze \cite{Nakagaki:2000,Nakagaki:2004}, creating efficient and fault-tolerant networks \cite{Tero:2010} and solving the travelling salesman problem \cite{Zhu:2018}. Algorithms inspired by Physarum polycephalum have been implemented for various optimization problems, including network design for multi-commodity flows \cite{Gao:2019,Li:2020,Ornek:2022}.

\begin{figure*}
		\includegraphics[width=1\textwidth]{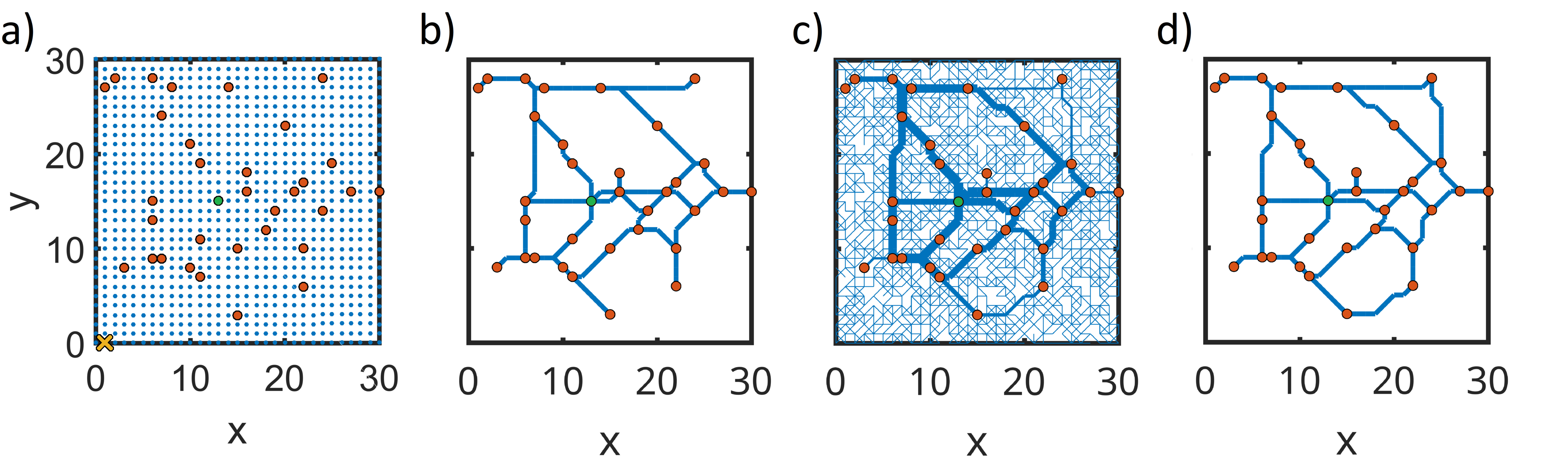}
		\caption{\label{Fig:1} Network self-organization for a multi-commodity flow problem in the presence of stochastic fluctuations. The sources and sinks simulate the metro stations of Tokyo railroad on a grid of 31 x 31 nodes with 528 demands, as illustrated in (a). The red dots represent the relative locations of the major cities around Tokyo, Tokyo itself is indicated by the green dot. The network design results from the dynamics of the edges, which are modelled by time-varying edge capacities (Eq.\,\eqref{Eq:D}). The emerging network topology is shown in (b) for the noiseless case ($\alpha=0$). Subplot (c) displays a single trajectory in the presence of noise ($\alpha=0.002$) and for a specific choice of the activation function (sigmoidal with exponent $n=1.6$). The widths of the edges are proportional to the corresponding capacities. The backbone of the network in (c) is extracted using a filtering procedure, the filtered network is shown in (d). In all simulations, we initially set all edge capacities equal to the value $D_{u, v}^0 = 0.5$. Furthermore, for each commodity $i$, we fix the potential $p_x^i = 0$ at the node indicated by the yellow cross {as a reference}. Further details on the numerical simulations are reported in the main text. } 		
\end{figure*}

Bio-inspired algorithms find successful applications for multi-commodity flow problems \cite{Tero:2010,Lonardi:2021,Bonifaci:2022,Folz:2023}. A prominent example is a city transportation network: Each commodity models the passengers traveling between two given stations. Physarum-inspired algorithms for solving multi-commodity flow problems typically tend to identify 
{a} network satisfying the constraints according to a rule that promotes transport along shared routes and inhibits it when the flow along one edge is below a chosen threshold \cite{Bonifaci:2022}. In Ref.~\cite{Lonardi:2022}, such an algorithm was used to find the optimal routing of passenger flows through the network of the Paris metro. This study includes a comparative assessment of different activation functions governing the edge dynamics: When the activation functions are power-laws of the flow, the exponent of the power-law determines the topological properties of the emerging networks, and the network topologies undergo a phase transition 
{between} tree-like and loopy topologies \cite{Banavar:2000}. 

{In the present study, we analyze the influence of noise on multi-commodity flow problems, where the edge capacity is a variable that depends on the current flowing across the edge through a nonlinear (activation) function \cite{Tero:2007}.} We assume that the edge capacity can undergo {stochastic} fluctuations and {examine the emerging network topologies} for three different functional behaviors of the activation function. {In particular,} we determine the characteristics of the emerging networks as a function of the noise amplitude and analyze their robustness, transport efficiency, and cost. {For our analysis, we choose the same case study as in Refs.~\cite{Tero:2010,Bonifaci:2022}, where the graph has a number of demands} geometrically arranged to {mimic} the relative locations of the major cities around Tokyo. {The graph and the demands are illustrated in {Fig.~\ref{Fig:1}(a)}. Examples of the emerging networks are provided in Figs.~\ref{Fig:1}(b) and (c).}

Our study extends recent work on optimization of transport in simple systems, consisting of either a single or two commodities, and in the presence of Gaussian white noise \cite{Meyer:2017,Folz:2021,Folz:2023}. These works showed that, for a finite range of noise amplitudes, the interplay between noise and a sigmoidal activation function can lead to the most robust solution in a relatively short convergence time. The phenomenology is reminiscent of noise-induced coherent effects, found in models simulating forest fires \cite{Meron:1992} and neurons \cite{Lindner:2004}, and include phenomena such as stochastic and coherence resonance \cite{Gammaitoni:1998,Lindner:2004,Perc:2007}, synchronization \cite{Nakao:2007,Boccaletti:2002}, and noise-induced phase transitions \cite{VandenBroeck:1994,Sagues:2007}. 

Our research question, the role of the specific activation function on network self-organization is {also} motivated by the observation that, 
in deep learning, the choice of the activation function determines the network expressivity \cite{Bahri:2020,Guehring:2020}. Within a different, yet connected, framework, here we study how noise-induced network self-organization depends on the choice of one of three representative functionals, which have been considered in the literature of multi-commodity transport.  

This paper is organized as follows. In Sec.\,\ref{Sec:transport}, we introduce the model {and} define the measures used to evaluate the emerging network topologies {for a graph with demands mimicking the Tokyo railroad transport problem}. 
We then determine the network measures as a function of the demand strength for the three different activation functions with no noise. Section \ref{Sec:stochastic} is devoted to the numerical methods and the algorithms used to extract the network topology from the stochastic dynamics. The network measures as a function of the noise amplitude are presented in Sec.\,\ref{Sec:results}. Conclusions and outlook are drawn in Sec.\,\ref{Sec:conclusions}.

\section{Noiseless transport on a grid}
\label{Sec:transport}

In this section, we introduce the model, define the network measures used for evaluating the topologies {and apply the multi-commodity flow problem to routing transport demands with the geometry of the greater area of Tokyo.} We focus on noiseless transport and study the network topologies for different activation functions.

\subsection{The model}
\label{Sec:model}

\begin{table*}[]
    \centering
  \begin{tabular}{ | c | c | c |}
    \hline
     \textbf{Activation function} & \textbf{Functional form} & \textbf{Threshold} \\ \hline
     Two-norm & $f(\{Q_{u, v}^{(i)}\}) = \sqrt{(Q_{u, v}^{(1)})^2 + ... + (Q_{u, v}^{(k)})^2}$ & - \\ \hline
     Hill/sigmoidal & $f(\{Q_{u, v}^{(i)}\}) = (Q_{u, v})^n/(\kappa^n + (Q_{u, v})^n)$ & $\kappa = 1$ \\ \hline
     ReLU & $f(\{Q_{u, v}^{(i)}\}) = (Q_{u, v} - \kappa) \theta(Q_{u, v} - \kappa)$ & $\kappa = 0.01$ \\
    \hline
  \end{tabular}
    \caption{
    {Three classes of activation functions.} Here, $Q_{u, v}$ denotes the total flow along the edge, which is determined as $Q_{u, v}=\sum_i|Q_{u, v}^{(i)}|$ (one norm). For the Hill function, we consider $n=1.6$ and $n=2$. In the definition of the ReLU function, $\theta(x)$ is Heaviside's function ($\theta(x)=1$ for $x>0$ and $\theta(x)=0$ otherwise).}
    \label{tab:activation}
\end{table*}




{The multi-commodity flow problem is represented by a set of $k$ transport demands on a graph that discretizes the space and is composed of $N$ nodes. 
Each node, labelled $u=1,\ldots,N$, is connected to {a number of other nodes}, described by the set $E_u$. We denote the edge connecting nodes $u$ and the neighbor $v\in E_u$ by the pair $(u,v)$. The edge length is {given by} $L_{u,v}$.} 

{Each demand is formed by a pair of a source node $s_+^i$, where a flow is injected ($+I^i$), and a sink node ($s_-^i$), where it is extracted ($-I^i$).} The demand $i$ is realized by a flow $I^i$ across the edges of the network. The flow through the edge connecting nodes $u$ and $v$ is denoted as $Q_{u, v}^i$. 
At the sources and sinks, it obeys the constraints:
\begin{equation}
\label{Eq:source-sink}
    \sum_{v \in E_{s_\pm^i}} Q_{s_\pm^i, v}^i=\pm I^i\,.
\end{equation} 
At the other nodes, instead, the flow is conserved:
\begin{equation}
\label{Eq:transit}
   \sum_{v \in E_u} Q_{u, v}^i=0\,. 
\end{equation}
For an electrical circuit, this is Kirchhoff's law and $Q$ is the electrical current. In the dynamics of network self-organization, these nodes are decision points. The flow of commodity $i$ is directed along the edges $(u,v)$ with non-vanishing edge capacity $D_{u, v}(t)$ and obeys the equation
\begin{equation}
\label{Eq:Q}
    Q_{u, v}^i(t) = \frac{D_{u, v}(t)}{L_{u, v}} (p_u^i(t) - p_v^i(t))\,,
\end{equation}
where $p_u^i$ is a potential (or pressure) at the node $u$ for commodity $i$. In an electrical circuit, which is the example we will refer to throughout this paper, Eq.\,\eqref{Eq:Q} is Ohm's law and $D_{u, v}(t)$ is a dynamical conductivity. The dynamics of the variables $p_v^i$ and $D_{u,v}$ determine the resulting network.

\subsection{The potential}

The potential $p_u^i$ is determined for each demand $i$ as a function of $D_{u, v}(t)$ by solving the linear set of equations in Eq.\,\eqref{Eq:Q}{, subject to the boundary conditions at the source and sink nodes given by Eq.\,\eqref{Eq:source-sink}, and Kirchhoff's law at any other node, Eq.\,\eqref{Eq:transit}}. {We fix the potential $p_x^i = 0$ at the node indicated by the yellow cross in Fig.\,\ref{Fig:1}(a) as a reference}. Let $N$ be the number of network nodes and {$b^i \in \mathbb{R}^N$} the vector representing the constraints, such that it has a value of $+I^i$ at the source node $s_+^i$ and $-I^i$ at the sink $s_-^i$. Otherwise, the entry is zero. {Then,} the vector $p^i = (p_1^i,\ldots, p_N^i)$ {containing} all node potentials associated to demand $i$ is determined by the linear system of equations \cite{Bonifaci:2022, Lonardi:2022}:
\begin{align}
\label{Eq:A}
    M p^i = b^i\,,
\end{align}
where $M$ is a $N\times N$ matrix,  
\begin{equation}
\label{Eq:M}
    M = ADL^{-1}A^T\,,
\end{equation}
with: (i) $D = \text{diag}(D_{e_1}, ..., D_{e_m})$ the $m\times m$ diagonal matrix whose diagonal elements are the edge capacities (conductivities), (ii) $L = \text{diag}(L_{e_1}, ..., L_{e_m})$ the $m\times m$ diagonal matrix whose eigenvalues are the edge lengths, and (iii) $A$ the $N\times m$ node-arc incidence matrix of the network, i.e., the column $(A^T)_e$ has a value of 1 in position $u$ and a value of -1 in position $v$ for all edges $e = (u, v)$. 

\subsection{The edge capacity (conductivity)}

The edge capacity (conductivity) $D_{u, v}(t)$ obeys the nonlinear differential equation \cite{Meyer:2017,Folz:2021}
\begin{equation}
\label{Eq:D:0}
    \partial_t D_{u, v} = f(\{Q_{u,v}^{(i)}\}) - \gamma D_{u, v}\,,
\end{equation}
where $f(\{Q_{u,v}^{(i)}\})$ is the activation function modelling the feedback of the flows associated to the commodities $j=1,\ldots,k$ on the evolution of the edge conductivity. 
The edge capacities are damped with rate $\gamma$ and hence edges with little flow on them (in particular, flow below a fixed threshold) are suppressed.

The network topology is determined by the resulting edge capacities $D_{u,v}$. They are found by integrating Eq.\,\eqref{Eq:D:0} using Eq.\,\eqref{Eq:Q}, where the potential is found from {Eqs.\,\eqref{Eq:A} and \eqref{Eq:M}}.

\subsection{The activation function}

\begin{figure*}
		\includegraphics[width=1\textwidth]{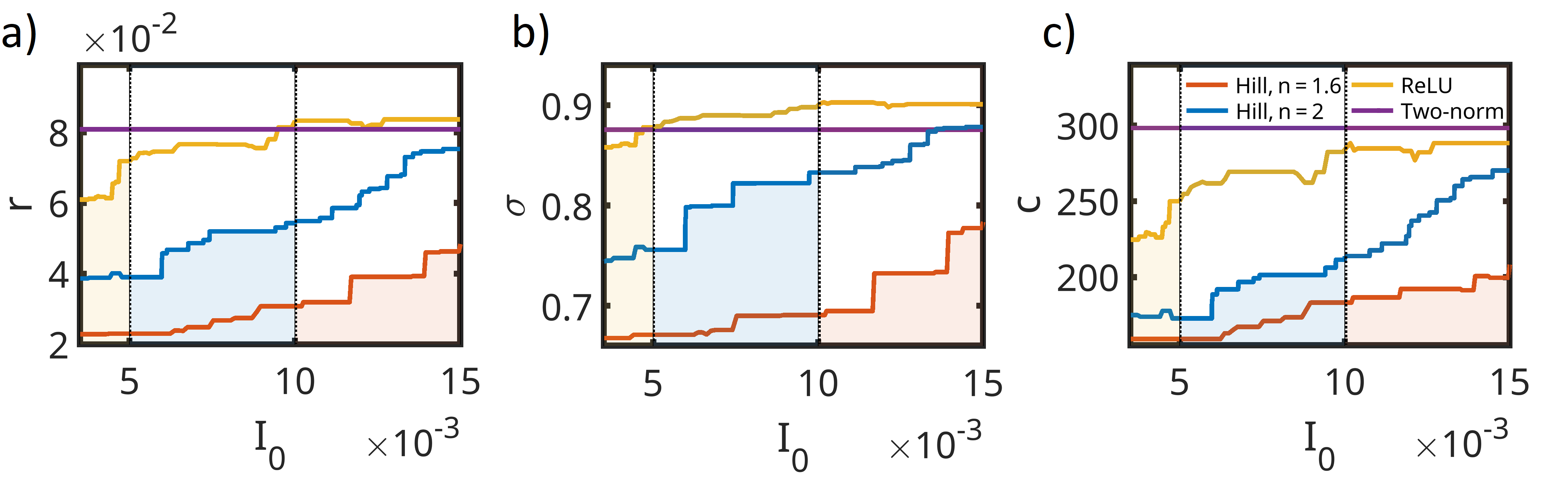}
		\caption{\label{Fig:2} Network measures for the noiseless dynamics and as a function of the demand strength $I_0$. The subplot display (a) the robustness $r$, (b) the transport efficiency $\sigma$ and (c) the cost $c$. They are extracted from the network obtained by integrating the coupled equations {\eqref{Eq:Q} - \eqref{Eq:D:0}} for a time of $1000\,\gamma^{-1}$ using a different activation function in each case, see legenda and Tab.\,\ref{tab:activation}. The measures of the networks are calculated after applying the disparity filter, which we detail in Sec.\,\ref{Sec:disparity}.
  In the rest of this work, we will focus on the range of $I_0$ indicated by the shaded regions. 	}	
\end{figure*}

{We} consider three different classes of activation functions: the two-norm, the Hill (or sigmoidal) function, and the ReLU function. Each of them depends on the total flow across an edge, as shown in Tab.~\ref{tab:activation}. 
{
As a consequence, sharing of edges between different commodities is rewarded. The different functional forms of the activation functions leads to different behavior as we discuss next.}

{Of} the three activation functions, only the two-norm function is a homogeneous function of the flow. Specifically, it is a homogeneous function with degree 1. As a consequence, the dynamics is independent of the demand strength $I_0$. In fact, equations \eqref{Eq:D:0} and \eqref{Eq:Q} are invariant after rescaling the conductivity and the flow by $I_0$.

The ReLU function is not a homogeneous function (even though it tends to behave as it were a homogeneous function for demand strengths much larger than the threshold, $I_0\gg \kappa$). In this work, we consider values of $I_0< \kappa$, i.e., smaller than the threshold, where the response depends on the choice of $I_0$.

The sigmoidal function saturates: This limits the flow that can be transported through the edge. As the flow between two nodes approaches the maximum capacity of the edge, the dynamics tends to construct multiple connections between them. In the following, we consider two different powers $n$ for the Hill function and analyze the regime where the dynamics is sensitive to small gradients as the demand strength $I_0$ increases above threshold. 

The dynamics governed by the two-norm function has been extensively studied in Ref.~\cite{Bonifaci:2022}. 
{In that work, a Lyapunov function was determined, the limit of the dynamics was formally characterized, and it was shown that the limit optimizes a mixture of transport efficiency and network cost. } 
{No} Lyapunov functions for the Hill and the ReLU function are known. 
{Moreover, since the sigmoidal and the ReLU function are nonlinear and non-homogeneous functions of the flow,} it is generally difficult to systematically choose the parameters in order to compare the network topologies for different activation functions. For this reason, we fix the parameters using a phenomenological approach that takes as a reference the robustness of the network when the activation function is the two-norm function. We then choose the threshold $\kappa$ of the Hill and the ReLU function and identify the interval of values of the demand strength $I_0$ that give a comparable network robustness. 

\subsection{Performance measures of networks}
\label{Sec:performance}

In what follows, we characterize the resulting network as a function of $I_0$. For this purpose, we introduce measures for (i) the robustness, (ii) the transport efficiency and (iii) the cost of the network.

The robustness is measured by the inverse of the {average} effective resistance of the network \cite{Ellens:2013}:
\begin{equation}
\label{Eq:r}
    r = \frac{k}{\sum_{i=1}^k R^k}\,,
\end{equation}
where $$R^i = (p_{s_+^i}^i - p_{s_-^i}^i)/I^i$$ is the effective resistance between the source node $s_+^i$ and the sink node $s_-^i$ of demand $i$. The quantity $R^i$ takes into account both the number of connections between $s_+^i$ and $s_-^i$ as well as their lengths. The measure $r$ is an indicator of robustness against edge failure. In fact, adding edges and/or reducing the length of connections increases the measure~\cite{Ellens:2011}. This is a different, nevertheless similar approach to existing measures, where the fault-tolerance of a network is measured by counting the number of edges that can be removed without separating the network into two parts for example \cite{Tero:2010}. Note that we set all non-vanishing edge conductivities to the same value (here, $D_{u, v} = 0.5$) before calculating $r$, corresponding to a fixed resistance per length.

The transport efficiency $\sigma$ of the network is 
{defined as} the average over all demands of the length of the shortest path $d^i$ connecting the source and the sink node $s_+^i$ and $s_-^i$. It reads
\begin{equation}
\label{Eq:sigma}
  \sigma =\frac{k}{\sum_{i=1}^k d^i}\,.  
\end{equation}

The cost of the network $c$ is the total length, found by summing over the segments $L_{u, v}$ where the conductivity is non-zero: 
\begin{equation}
\label{Eq:c}
    c= \sum_{(u, v) \in E} L_{u, v}\,,
\end{equation}
with $E$ the set of all edges with $D_{u, v} > 0$.

These quantities are displayed in Fig.\,\ref{Fig:2} as a function of the demand strength $I_0$; the colors represent the different activation functions from Table \ref{tab:activation}. 

\begin{figure*}
		\includegraphics[width=1\textwidth]{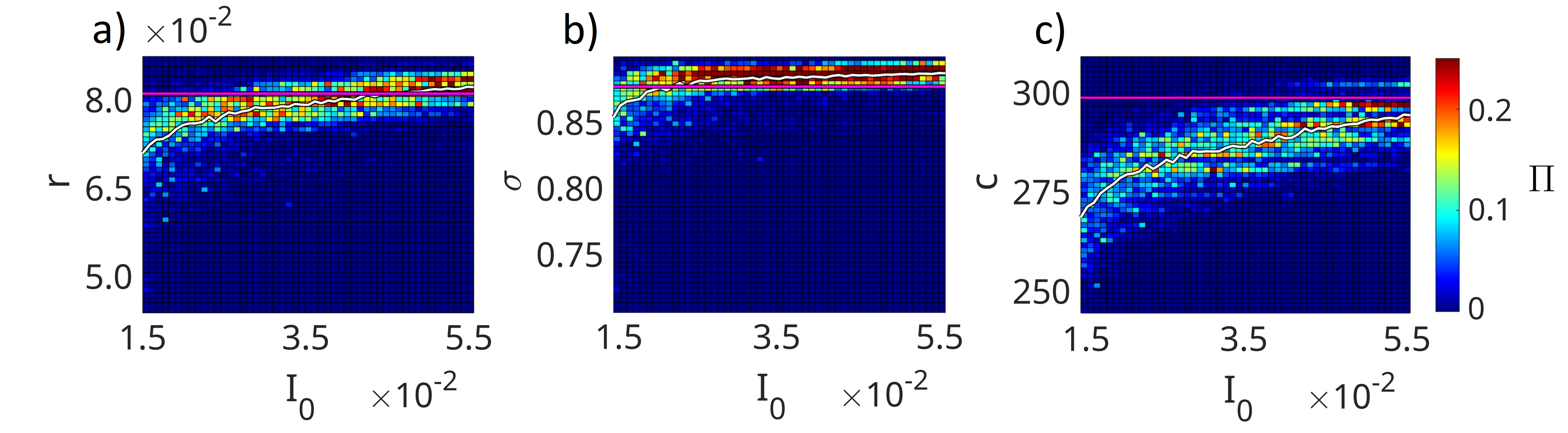}
		\caption{\label{Fig:4} Network measures as a function of the demand strength $I_0$ for the two-norm function and in the presence of Gaussian noise ($\alpha=0.001$). The panel displays (a) the robustness $r$, (b) the transport efficiency $\sigma$, and (c) the cost $c$. The color scale at a given value of $I_0$ indicates the percentage {$\Pi$} of 250 trajectories at the corresponding value of the measure (dark blue is statistically irrelevant, dark red corresponds to more than 20\%). The solid white line indicates the average for each value of $I_0$. The solid purple line shows the deterministic value ($\alpha=0$) for comparison. The network measures are calculated after applying the disparity filter (see text). Details on the numerical simulations are reported in Sec.\,\ref{Sec:numerical-methods}.}   
\end{figure*}

For the two-norm function, the performance of the networks is, as expected, independent of $I_0$. For the sigmoidal and the ReLU functions, the measures increase in a step-like fashion, indicating a discontinuous transition to different topologies of increasing robustness, transport efficiency, and cost. Each topology exists for a certain range of the demand strength $I_0$. For the ReLU function, the measures reach a constant value for $I_0 \gg \kappa$, whereas for the sigmoidal function, they keep increasing with $I_0$ due to the saturating character of the activation function that limits the maximum edge capacity. The measures can locally decrease for increasing $I_0$, which may be an artifact of our filtering procedure (see next Section and Appendix \ref{App:A}).

\subsection{{Application to the Tokyo transport problem}}
\label{Sec:transport-problem}

{We apply the model introduced in this section to the multi-commodity flow problem} represented by the distribution of sources and sinks illustrated by the red circles in Fig.\,\ref{Fig:1}(a). {We use the same setup} as in Refs.~\cite{Tero:2010,Bonifaci:2022}, where the sources and sinks {are geometrically arranged} to mimic the relative locations of the major cities around Tokyo. {A transport demand between each pair of cities is assumed, which means} that the emerging network has to satisfy $k = 528$ demands. 

{We consider} a grid of $N=31 \times 31$ nodes, of which 33 are the cities. Each node, labelled $u=1,\ldots,N$, is connected to a number of nearest and next-nearest neighbors. The edge length $L_{u,v}$ is set to unity when the nodes $u$ and $v$ are nearest neighbors and to $\sqrt{2}$ when the nodes are connected by a diagonal.

We choose $I^i = I_0$ for all demands $i$ except for those that involve the node at the relative location of Tokyo: For these demands, as in Ref.~\cite{Tero:2010}, we set $I^i = 7\,I_0$ to reflect the importance of Tokyo as the center of the region.

\section{Model and numerical methods for the stochastic dynamics}
\label{Sec:stochastic}

We finally come to the core of the paper, the study of the networks emerging from the interplay of stochastic dynamics and the nonlinear activation functions. The conductivity now evolves according to the equation
\begin{equation}
\label{Eq:D}
    \partial_t D_{u, v} = f(\{Q_{u,v}^{(i)}\}) - \gamma D_{u, v}+\sqrt{\gamma}\alpha\xi_{u, v}(t)
\end{equation}
with the stochastic force $\xi_{u, v}(t)$, whose amplitude is scaled by the parameter $\alpha$. The force is statistically defined by the average over an ensemble of trajectories: it has no net drift, $\langle \xi_{u, v} (t) \rangle = 0 $, and simulates Gaussian white noise with no spatial correlations, $\langle \xi_{u, v}(t) \xi_{u', v'}(t') \rangle = \delta_{u, u'} \delta_{v, v'} \delta(t - t')$ \cite{vanKampen:2007}. The network dynamics results from integrating the coupled equations \eqref{Eq:Q}, {\eqref{Eq:A}}, \eqref{Eq:M} and \eqref{Eq:D}. 

In what follows, we first introduce the numerical methods 
{used} to integrate the stochastic differential equations. Since the Langevin force $\xi_{u,v}$ gives rise to fluctuations of the edge conductivities, the stationary values of the measures have a finite variance. In order to eliminate the background noise, we apply the filter of Ref.\,\cite{Folz:2023} that allows us to identify the statistically relevant edges. In the Appendix \Ref{App:A}, we benchmark it with other filtering procedures. 

\begin{figure*}
		\includegraphics[width=1\textwidth]{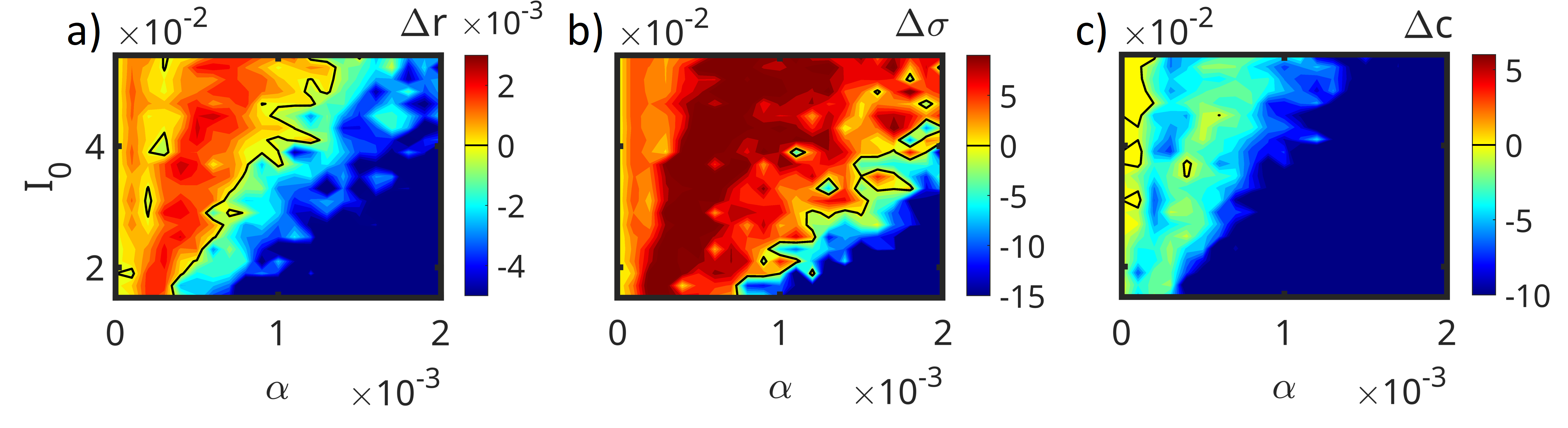}
		\caption{\label{Fig:6} Color plot of the difference between the measures of the stochastic and the noiseless networks as a function of $I_0$ and $\alpha$ for the two-norm activation function. The measures of the networks are evaluated after applying the disparity filter. Hot colors mean that the measure takes a larger value in the presence of noise, cold colors the opposite. Subplot (a) displays the difference $\Delta r = \langle r \rangle - r_0$ between the ensemble average $\langle r \rangle$ of the robustness and the value $r_0$ of the noiseless dynamics, (b) $\Delta \sigma = \langle \sigma \rangle - \sigma_0$ for the transport efficiency and (c) $\Delta c = \langle c \rangle - c_0$ for the cost. The mean value of the stochastic dynamics is calculated over 50 trajectories for each value of $I_0$ and $\alpha$. The solid black lines indicate the value where the difference exactly vanishes, the fluctuations are attributed to the relatively small number of trajectories.}   
\end{figure*}

\subsection{Numerical methods}
\label{Sec:numerical-methods}

The integration of Eq.\,\eqref{Eq:D} is performed using the Euler-Maruyama scheme, as outlined in Ref.\,\cite{Kloeden_1992}. We have analyzed convergence for different step sizes: For the parameter values considered here, we set the step size $\Delta t = 0.05\,\gamma^{-1}$. We then set the evolution time as $t_{\rm end}=1000\,\gamma^{-1}$. In fact, extensive tests over a statistically relevant ensemble of trajectories (namely, individual network evolutions) show that each trajectory has reached a (meta-)stable configuration.
In all simulations, the conductivities of all edges are initially equal to the value $D_{u, v}^0 = 0.5$. For each commodity $i$, we fix the potential $p_x^i = 0$ at the node indicated by the {yellow} cross in Fig.\,\ref{Fig:1}(a) {as a reference}. 

Figure \ref{Fig:1}(c) displays a typical network obtained {at time $t_{\rm end}$}. Edges with a non-zero conductivity $D_{u, v} > 0$ are drawn with blue lines, whose width is proportional to $D_{u, v}$. Noise leads to a fluctuating distribution of weak connections as well as to statistically relevant links, that are otherwise absent in the noiseless dynamics.
 
We remark that, as the reported topologies are obtained for finite integration times, we cannot claim that they are the steady state. The dynamics, in fact, is described by a multi-dimensional and nonlinear Fokker-Planck equation \cite{Frank:2005}, and it is not even guaranteed that a steady state exists. Nevertheless, over the considered time, the integrated trajectories converge relatively fast towards a certain cost, robustness, and transport efficiency and then remain stably trapped about these values, performing fluctuations of the order of the noise amplitude. To these configurations, we then apply the filtering procedure, which we describe in what follows.

\subsection{Filtering procedures}
\label{Sec:disparity}

For each simulation run, we apply a filtering procedure to the final conductivities $D_{u, v}$ to retrieve the backbone of the network. In order to identify the statistically relevant edges, we use the filter of Ref.\,\cite{Folz:2023}. This filter extends the method of Ref.\,\cite{Serrano:2009} to noisy networks. We discuss alternative filters in the appendix. These other filters lead to similar results giving evidence to the appropriateness of our filtering approach. 

At the end of our simulations, we average the conductivity over the time interval $\mathcal I_t=[t_{\rm end}-d_t,t_{\rm end}]$, where $t_{\rm end}$ is the total integration time and $d_t=1/\gamma$. The averaging levels out the fluctuations. We verified that the interval is long enough for averaging; in particular, changing $d_t$ by a factor 5 does not change the results.
We denote the time-averaged conductivities by $\langle D_{u,v}\rangle$: 
\begin{equation}
\langle D_{u,v}\rangle=\frac{1}{d_t}\int_{\mathcal I}{\rm d}\tau D_{u,v}(\tau)\,.    
\end{equation}
We then define $s_u = \sum_{v \in E_u} \langle D_{u, v}\rangle$ as the strength of node $u$ and normalize the conductivities of the incident edges as ${\mathcal P}_v=D_{u,v}/s_u$ such that $\sum_{v\in E_u} {\mathcal P}_v=1$. Hereby, $E_u$ denotes the set of edges that are connected to node $u$, and $v$ denotes the neighboring nodes. 

All edges whose conductivities are purely random are statistically not significant. Assume that $\langle D_{u, v}\rangle$ of an edge incident to node $u$ of degree $k$ is sampled from a random uniform distribution. To find the corresponding probability density $\rho(x)$, we use the method of induction. For $k=2$ edges, it holds ${\mathcal P}_1+{\mathcal P}_2=1$. We set ${\mathcal P}_1=x$ with a random number $x$ in the interval $[0, 1]$ and define the probability density $\rho(x)$ such that ${\mathcal P}_1=\rho(x) dx$. It follows that $\rho(x)=1$. For $k>2$, it holds $$\rho(x) dx= dx (k-1)\!\int_0^{1-x}dx_1\ldots\int_0^{1-x_{k-2}}dx_{k-3}\,,$$
yielding \cite{Serrano:2009}  
\begin{equation}
\label{Eq:rho}
    \rho(x)\, dx = (k - 1) (1 - x)^{k - 2} dx.
\end{equation}
The probability that the edge $(u, v)$ is compatible with the null hypothesis, i.e. is purely random, is given by \cite{Serrano:2009}
\begin{equation}
\label{Eq:disparity}
    \beta_{u, v} = 1 - (k - 1) \int_0^{D_{u, v}} (1 - x)^{k - 2} dx.
\end{equation}
This value shall be compared with a threshold $\beta$, the significance level, which we choose in the interval $\beta \in [0, 1]$. For $\beta_{u, v} \ge \beta$, the edge is statistically not significant and filtered out. 

The choice of the significance level has a certain arbitrariness (see, e.g., the discussion in Ref.\,\cite{Serrano:2009}). We reduce this arbitrariness by benchmarking our results with the results of other filtering procedures, as shown in Appendix \ref{App:A}. In the following, we use $\beta = 0.12$ for the two-norm function and $\beta = 0.3$ for all other activation functions, yielding comparable values of the network measures. 

The significance level limits the range of noise amplitudes for which the filter is useful. In fact, for sufficiently large values of $\alpha$, the stochastic dynamics becomes dominant so that it fails to extract the backbone of the network. We thus limit our analysis to noise amplitudes for which the filter can be successfully applied. 

\section{Network topologies in the presence of noise}
\label{Sec:results}

In this section, we present the network topologies emerging from the interplay of noise and nonlinear dynamics for each of the three classes of activation functions. 
We investigate the (filtered) networks as a function of the demand strength $I_0$ and the noise amplitude $\alpha$. For each fixed pair of $\alpha$ and $I_0$, we evaluate an ensemble of 250 trajectories and determine the robustness, the transport efficiency, and the cost of the backbone of each individual trajectory as defined in Sec.\,\ref{Sec:performance}. 

\subsection{Two-norm function.}

\begin{figure}
		\includegraphics[width=0.5\textwidth]{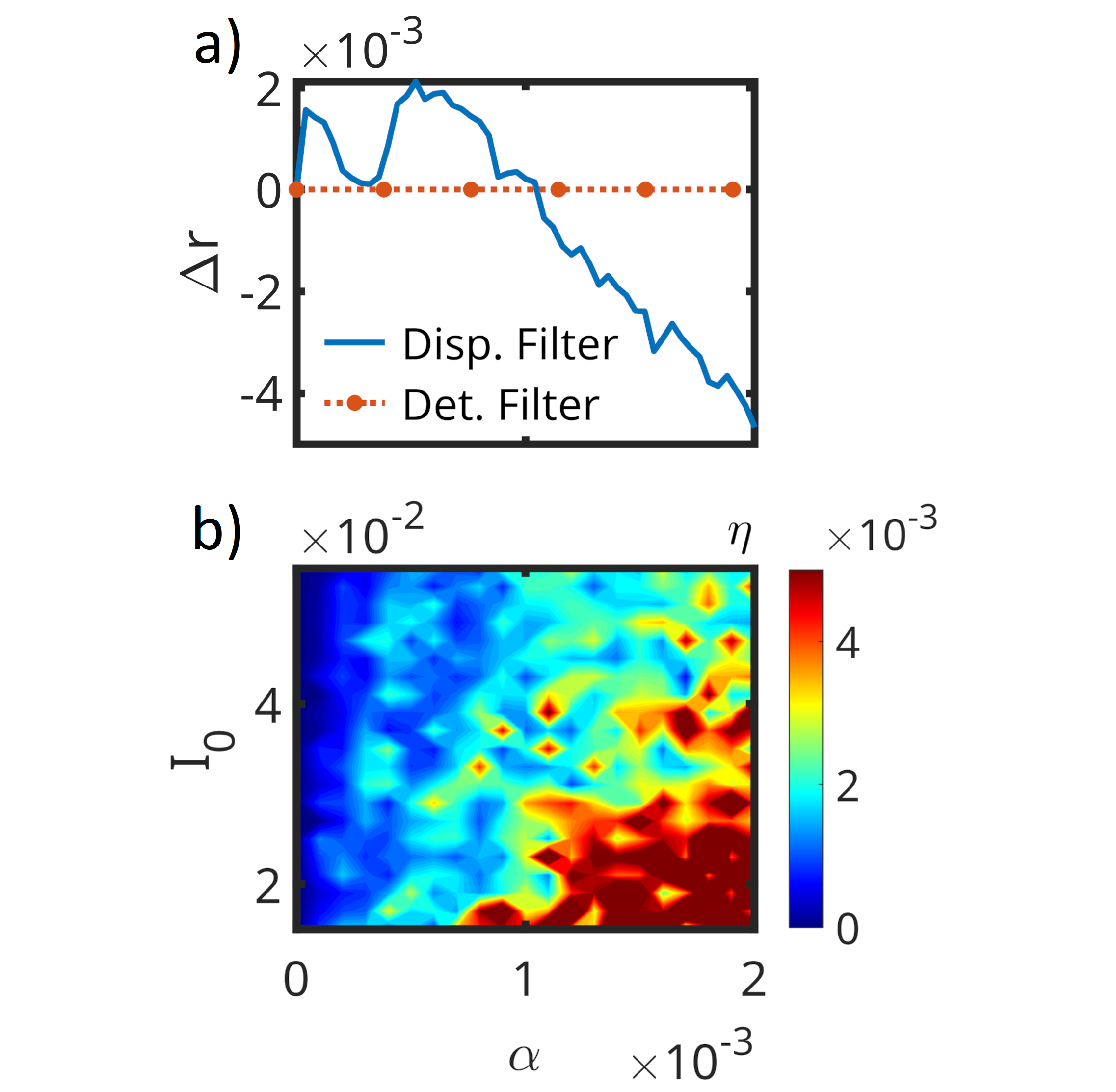}
		\caption{\label{Fig:12} {(a)} The solid line indicates the difference $\Delta r = \langle r \rangle - r_0$ between the ensemble average $\langle r \rangle$ of the robustness and the value $r_0$ of the noiseless dynamics as a function of the noise amplitude $\alpha$ for $I_0 = 4.5 \times 10^{-2}$. Each data point represents the average value over 250 simulation runs. The red dots indicate the difference $\Delta r$ after applying the deterministic filter to the noisy networks for a time $t = 2000\,\gamma^{-1}$. {(b)} The standard deviation of the robustness distribution shown in Fig.\,\ref{Fig:6} as a function of the demand strength $I_0$ and the noise amplitude $\alpha$.}   
\end{figure}

We first discuss the networks generated using the two-norm function for Gaussian, white noise. The white solid line in Fig.\,\ref{Fig:4} displays (a) the mean robustness, (b) the mean transport efficiency, and (c) the mean cost of the network as a function of $I_0\in [1.5 \times 10^{-2},\,5.5 \times 10^{-2}]$ and for a fixed noise amplitude $\alpha = 0.001$. The mean is taken over 250 trajectories and shows that noise tends to spread out the flow across the network. The trajectory distribution for each $I_0$ is indicated by the color scale: the measures of each trajectory cluster about the mean value. 

{We note} that the topology depends on $I_0$. This is in striking contrast with the noiseless behavior (magenta curves), where all measures are independent of the demand strength $I_0$. The average robustness, transport efficiency, and cost are generally increasing as a function for $I_0$. 
For small $I_0$ ($I_0\sim 1.5 \times 10^{-2}$), the mean values of the robustness, transport efficiency and cost are lower than in the noiseless case. For larger values of $I_0$ ($I_0\sim 5.5 \times 10^{-2}$), 
the mean value of robustness and transport efficiency slightly exceed the corresponding noiseless measures (see Figs.\,\ref{Fig:4}(a) and (b)), whereas the average cost is smaller. At values of $I_0$ outside the interval, these measures all saturate to their values in the noiseless case. {Remarkably}, there is a regime where the dynamics converges to more efficient network designs, namely, networks that are more robust and efficient, and at the same time less costly.

In order to identify the regimes where noise leads to more efficient topologies, we analyze the difference between the measures with and without noise over a broad range of demand strengths $I_0$ and noise amplitudes $\alpha$. Figure \ref{Fig:6} displays the differences $\Delta y=\langle y\rangle -y_0$ of the mean values over the noisy trajectories $\langle y\rangle$ from the noiseless measure $y_0$ for (a) robustness $y=r$, (b) transport efficiency $y=\sigma$ and (c) cost $y=c$. The differences are displayed as color plots in the $I_0-\alpha$ plane. We first note that the contours of the equipotential lines follow an underlying linear behavior $I_0\propto \alpha$. In fact, in the presence of noise, Eq.\,\eqref{Eq:D} regains the invariance by $I_0$ when rescaling $\alpha\to \alpha I_0$. The plot indicates that all measures tend to the noiseless case for $I_0\gg \alpha$ when the stochastic term becomes an infinitesimally small perturbation to the dynamics governed by Eq.\,\eqref{Eq:D}. 

\begin{figure*}
		\includegraphics[width=1\textwidth]{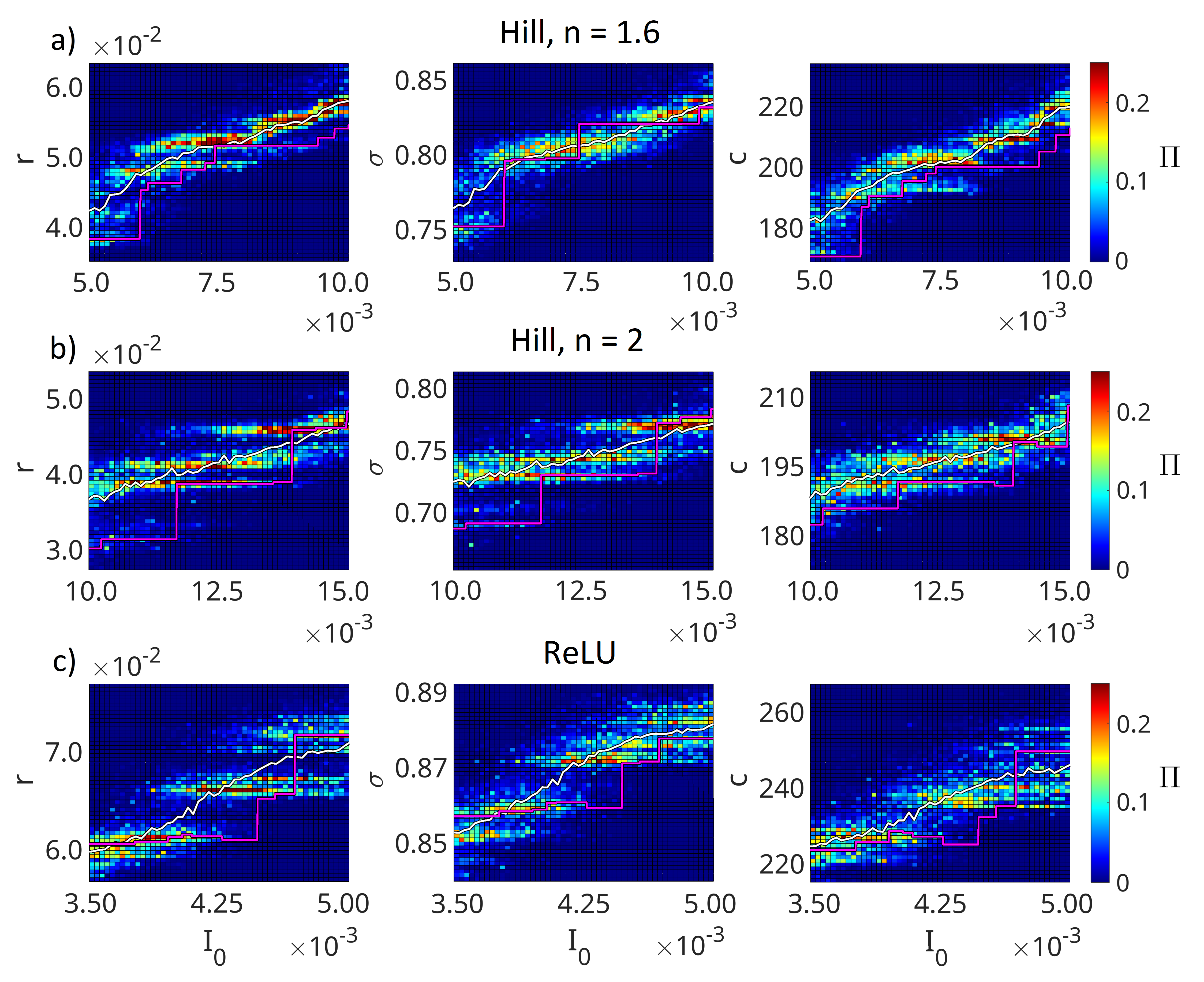}
		\caption{\label{Fig:8} Network measures as a function of the demand strength $I_0$ and for $\alpha=0.001$. The panels in row (a) are obtained using the sigmoidal function with $n = 1.6$, in row (b) using the sigmoidal function with $n = 2$, and in row (c) using the ReLU function. The panels display (left) the robustness $r$, (central) the transport efficiency $\sigma$ and (right) the cost $c$. The color scale at a given value of $I_0$ indicates the percentage {$\Pi$} of 250 trajectories at the corresponding value of $r$ (dark blue is statistically irrelevant, dark red corresponds to more than 20\%). The solid white line indicates the average for each value of $I_0$. The solid purple line shows the noiseless value and is plotted for comparison. The results of the noisy dynamics are invariant under the filtering procedure we apply. Details on the numerical simulations are reported in Sec.\,\ref{Sec:numerical-methods}.}
\end{figure*}

It is tempting to assume that noise will lead to inferior networks. This is true for large enough noise. However, small noise can be beneficial. In the dark red part of the diagram the noisy solutions are more robust and efficient than the noiseless networks and the network topologies reach the largest values of $r$ and $\sigma$.  A cut of the plot at a fixed $I_0$, Fig.~\ref{Fig:12}{(a)}, indicates a resonance-like behavior as a function of the noise amplitude $\alpha$.
Remarkably, in this region, the standard deviation of the robustness decreases, see Fig.\,\ref{Fig:12}{(b)}. This behavior challenges the conventional expectation that the size of the fluctuations increase with $\alpha$.  


Following the criterion introduced in Ref.\,\cite{Folz:2023}, we denote the region of increased robustness and efficiency as a noise-induced resonance. This terminology is borrowed from nonlinear dynamics, where noise-induced coherent effects have been reported in models simulating forest fires \cite{Meron:1992} and neurons \cite{Lindner:2004}. As in stochastic resonance \cite{Gammaitoni:1998,Lindner:2004,Perc:2007,Folz:2021}, we observe an optimal range of noise amplitudes at which noise increases the efficiency of the algorithm leading to more robust network topology than in the noiseless case. We note that these solutions are not fixed points of the noiseless dynamics, as we verify by taking them as initial conditions and determine the evolution according to Eq.\,\eqref{Eq:D:0}. This procedure, which we refer to as the deterministic filter and detail in Appendix \ref{App:A}, shows that for sufficiently long integration times, $\Delta r\to 0$, as demonstrated by the red dots in Fig.\,\ref{Fig:12}{(a)}. This suggests that the topologies obtained using the noisy dynamics are novel solutions, which are absent in the noiseless case.


\begin{figure*}
		\includegraphics[width=1\textwidth]{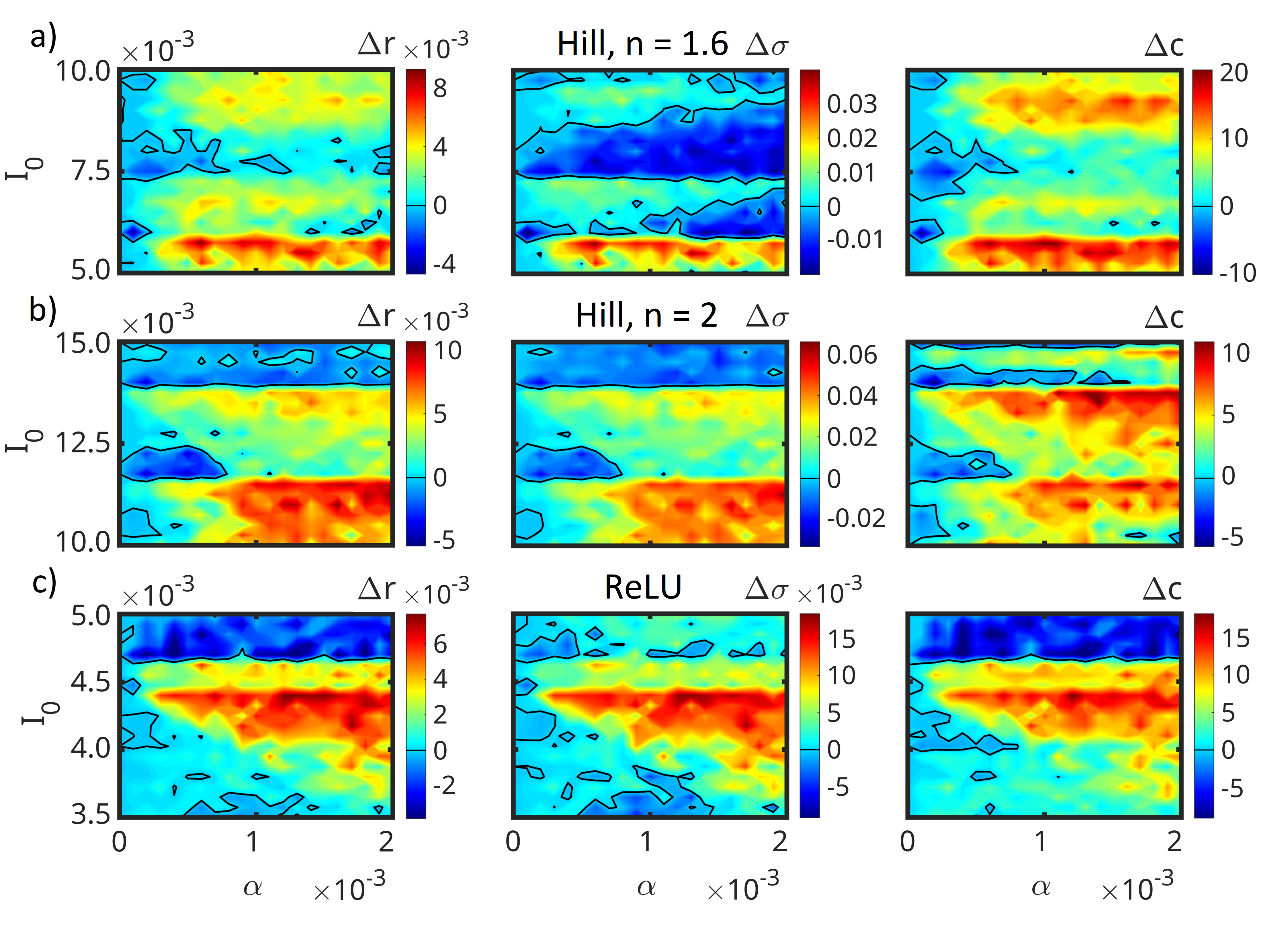}
		\caption{\label{Fig:10} Comparison between the measures of the stochastic and the noiseless networks as a function of $I_0$ and $\alpha$ when the activation function is (row a) the Hill function with $n = 1.6$, (row b) the Hill function with $n = 2$ and (row c) the ReLU function. The panels display (left) the robustness difference $\Delta r = \langle r \rangle - r_0$, (central) the transport efficiency difference $\Delta \sigma = \langle \sigma \rangle - \sigma_0$ and (right) the cost difference $\Delta c = \langle c \rangle - c_0$ between the corresponding ensemble averages $\langle r \rangle, \langle \sigma \rangle, \langle c \rangle$ of the stochastic case and the values $r_0, \sigma_0, c_0$ of the deterministic case. The mean value of the stochastic dynamics is calculated over 50 trajectories for each value of $I_0$ and $\alpha$. The solid black lines indicate the value where the difference exactly vanishes. The fluctuations in these lines are attributed to the relatively small number of trajectories.}   
\end{figure*}

\subsection{Sigmoidal and ReLU functions.}

We now turn to the sigmoidal (Hill) and the ReLU function. Figure \ref{Fig:8} displays the network measures as a function of $I_0$ and for fixed, non-vanishing noise amplitude $\alpha$ for three cases: in row (a) we report the network measures for the sigmoidal function with exponent $n=1.6$, in row (b) for the sigmoidal function with exponent $n=2$, and in row (c) for the ReLU function. 

The behavior of the mean value as a function of the demand strength (white solid line) indicates that  the measures generally increase with $I_0$. The comparison with the noiseless measures (magenta curves) indicates the existence of multiple intervals of $I_0$ where noise leads to more robust networks. These networks simultaneously optimize transport, at the expense of larger cost. At each $I_0$, the trajectories are clustered about the fixed points of the noiseless case: The effect of noise is to increase the range of stability of certain topologies, extending the plateaus to lower values of $I_0$ (compare with Fig.\,\ref{Fig:2}). 

\begin{figure*}
		\includegraphics[width=1\textwidth]{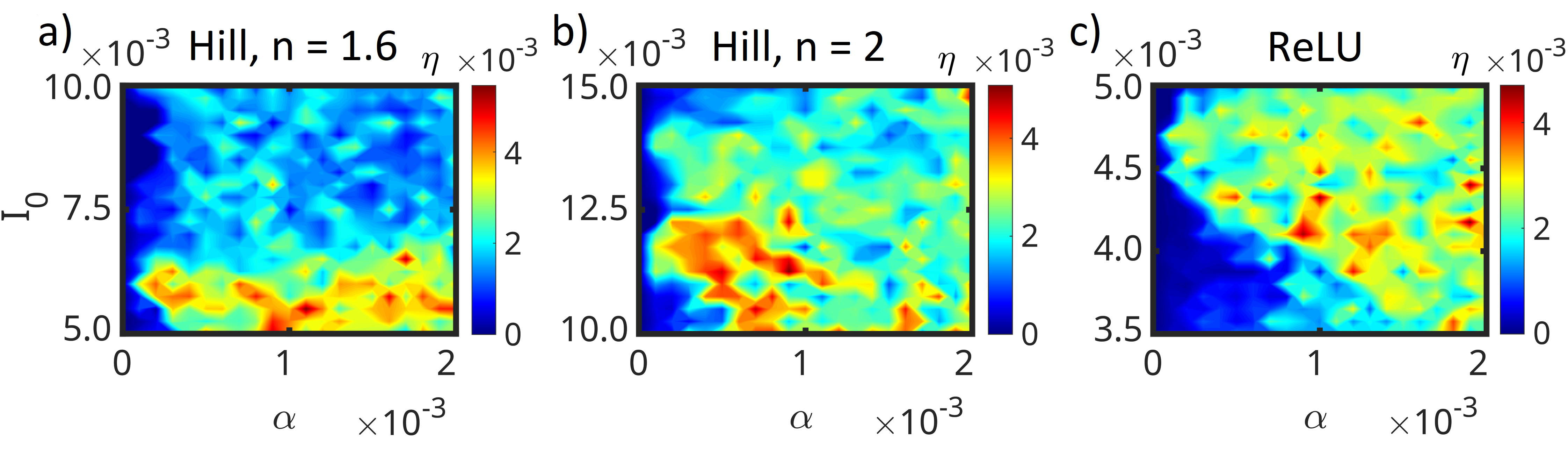}
		\caption{\label{Fig:11} The standard deviations of the robustness distributions shown in Fig.\,\ref{Fig:10} as a function of the demand strength $I_0$ and the noise amplitude $\alpha$. The subplots correspond to the different activation functions: (a) the sigmoidal with $n = 1.6$, (b) the sigmoidal with $n = 2$, and (c) the ReLU.}   
\end{figure*}

The contour plots in Fig.\,\ref{Fig:10} show the regions in the $I_0-\alpha$ plane where noise leads to network topologies with larger robustness, transport efficiency and cost (see dark red colored regions). 
For the sigmoidal function with $n=2$ and the ReLU function, the size of these regions increases monotonously with $\alpha$. 
Figure \ref{Fig:11} illustrates the standard deviation of the corresponding robustness distribution: the variance does not seem to depend on $\alpha$. A more detailed analysis actually shows that the standard deviation narrows in correspondence of the regions where noise leads to more robust networks. This indicates a resonance-like response to noise.
We verified that the topologies found by the noisy dynamics do not depend on the filtering procedure we apply, including the deterministic filter. This shows that they are also solutions of the noiseless dynamics. Noise, in this case, modifies the landscape of local minima by favoring the more robust solutions.

In summary, we find that also for the sigmoidal and ReLU functions, noise can be used for optimizing robustness and efficiency of the emerging network topologies. This underscores the generality of our results, indicating that noise can enhance optimization algorithms for different classes of nonlinearities.

\section{Conclusions}
\label{Sec:conclusions}

We have {investigated} the network topologies for a multi-commodity flow problem and compared the robustness, transport efficiency, and cost of the emerging network topologies in the presence and in the absence of stochastic fluctuations. The networks have been calculated on a graph with a geometry of demands representing the relative locations of the major cities around Tokyo. The equations used for determining the topology consist of evolving the edge capacity according to Eq.\,\eqref{Eq:D}. The growth of the edge capacity depends nonlinearly on the flow along the edge according to an activation function. Moreover, we have assumed that the edge capacity can additionally experience stochastic fluctuations. 

We have analyzed different nonlinear functions of the flow and determined the behavior of the measures of the resulting self-organized networks as a function of the demand strength and the noise amplitude. The response to noise is different, yet for all the considered activation functions, we can identify parameter regimes where the interplay of noise and the nonlinear activation function gives rise to a resonance-like convergence to a more robust network topology. For the specific case of the two-norm function, the noisy dynamics converges to solutions that are otherwise unstable in the absence of noise. For the sigmoidal and the ReLU, instead, noise favors the most robust solutions of the deterministic dynamics. These results support and complement previous insights obtained with a simplified multi-commodity flow problem consisting of two demands \cite{Folz:2023} and demonstrates their generality. It indicates that noise can be a resource for optimization algorithms. 

Interestingly, the features of the noise-induced resonances can change dramatically depending on the activation function. This observation suggests a different strategy for optimization, such that for a given noise and demand strength, one could identify a class of activation functions that maximizes the robustness and the transport efficiency of the network. Remarkably, the activation functions of this dynamics play a similar role in neural networks, where it is key for the expressivity \cite{Guehring:2020}. In general, this study puts forward the need of identifying an adequate functional that permits one to systematically shed light on the interplay of nonlinearity and stochastic dynamics for optimization.

\acknowledgements
    
The authors thank Philipp H\"ovel for stimulating discussions and acknowledge support from the Deutsche Forschungsgemeinschaft (DFG, German Research Foundation) Project-ID No.429529648, TRR 306 QuCoLiMa (Quantum Cooperativity of Light and Matter), from the Bundesministerium f\"ur Bildung und Forschung (BMBF, German Ministry of Education and Research) under the grant "NiQ: Noise in Quantum Algorithms", and in part by the National Science Foundation under Grants No. NSF PHY-1748958 and PHY-2309135. 
	
\begin{figure*}
		\includegraphics[width=1\textwidth]{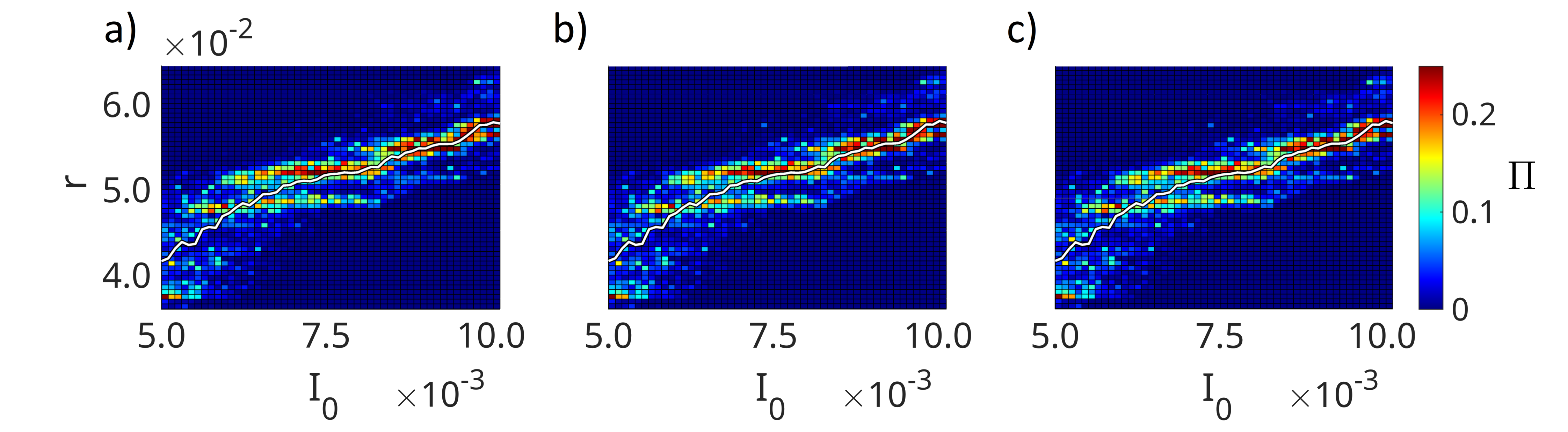}
		\caption{\label{Fig:20} The robustness $r$ as a function of the demand strength $I_0$ for the case of using the filtering procedure based on (a) the disparity filter, (b) a cut-off filter with threshold $d = 0.05$, and (c) the deterministic filter, applied for a time $t = 1000\,\gamma^{-1}$. For each value of $I_0$, a total of 250 simulation runs were performed. The solid white line represents the ensemble average, and the color code indicates the probability distribution: dark blue is statistically irrelevant, and dark red corresponds to more than $20\,\%$ of the simulation runs, yielding a network topology with the respective robustness. For all simulations, we chose $\alpha = 10^{-3}$. Details on the filtering procedures are given in the main text.}   
\end{figure*}

\begin{appendix}
\section{Testing the disparity filter}
\label{App:A}

We benchmark the disparity filter using two different procedures. In all cases, we first divide the time-averaged conductivity by the maximal edge conductivity, $D_{\rm max}=\underset{(u, v) \in E_0}{\max} \langle D_{u, v} \rangle$, where $E_0$ is the set of all edges:
$$D_{u, v} \to D_{u,v}'\equiv\langle D_{u, v} \rangle/D_{\rm max} \,.$$

\noindent {\it Disparity filter:} We first apply a constant offset $\delta_1$, such that 
$$D_{u,v}' \to D_{u,v}' + \delta_1\,.$$ 
We then apply the disparity filter as described in Sec.\,\ref{Sec:disparity}. The offset is introduced in order to level out fluctuations left after the time average. It is necessary to avoid that the filtering procedure artificially amplifies fluctuations due to noise. For the given choice of $d_t$ and for the considered values of $\alpha$, we take $\delta_1=0.05$. 


\noindent {\it Cutoff filter:} We consider the threshold $\delta_2= 0.05$ and set to zero all conductivities with $D_{u,v}'<\delta_2$. The resulting network is composed of all non-vanishing edges.

\noindent {\it "Deterministic" filter:} We take $D_{u,v}'$ as initial conditions of a numerical integration according to the noiseless evolution of Eq.\,\eqref{Eq:D:0} and then verify whether it converges to the filtered network over the same integration time $t_{\rm end}$.

In all cases, we remove any dead ends after the filter was applied. We show the comparative assessment for the robustness using the Hill function with $n = 1.6$. Figure \Ref{Fig:20} displays the robustness $r$ as a function of the demand strength $I_0$ for the three different filtering procedures. We note that the qualitative behavior is the same for all filtering procedures, indicating that our results are independent of the specific choice of the filtering procedure.

\end{appendix}

\bibliography{biblio-Physarum.bib} 
\end{document}